\documentclass[aps,eqsecnum]{revtex4}
\usepackage{epsfig}
\usepackage{latexsym}
\usepackage{bm}

\begin{document}
%======================================%
%<<<<<<<<<<<< TITLE PAGE >>>>>>>>>>>>>>%
%======================================%
\thispagestyle{empty}

%
% Preprint numbers
%
{\baselineskip0pt
\rightline{\large\baselineskip14pt\rm\vbox
        to20pt{\hbox{OCU-PHYS-209}
               \hbox{AP-GR-14}
               \hbox{OU-TAP-228}
               \hbox{YITP-04-14}
\vss}}
}
\vskip15mm

\begin{center}

{\Large\bf An Improved Search Method for Gravitational Ringing of Black Holes}

\bigskip

Hiroyuki Nakano$^1$, Hirotaka Takahashi$^{2,3,4}$, 
Hideyuki Tagoshi$^{3,5}$ and Misao Sasaki$^4$

\smallskip
$^1${\em Department of Mathematics and Physics,~Graduate School of 
 Science,~Osaka City University,\\ Osaka 558-8585, Japan
}

\smallskip
$^2${\em Department of Physics,~Graduate School of Science and Technology,
Niigata University,\\ Niigata 950-2181, Japan}\\

\smallskip
$^3${\em Department of Earth and Space Science,~Graduate School of 
 Science,~Osaka University,\\ Toyonaka, 560-0043, Japan
}

\smallskip
$^4${\em Yukawa Institute for Theoretical Physics,
Kyoto University,\\ Kyoto 606-8502, Japan}\\

\smallskip
$^5${\em Theoretical Astrophysics, 
California Institute of Technology, 
Pasadena, CA 91125, USA}\\
\smallskip

\medskip

\today

\end{center}

\bigskip

A black hole has characteristic quasi-normal modes 
that will be excited when it is formed or when the geometry is
perturbed. The state of a black hole when the quasi-normal modes 
are excited is called the gravitational ringing, 
and detections of it will be a direct
confirmation of the existence of black holes.
To detect it, a method based on matched filtering needs to be
developed. Generically, matched filtering requires a large 
number of templates, because one has to ensure a proper match
of a real gravitational wave
with one of template waveforms to keep the detection efficiency
as high as possible. On the other hand, the number of templates must
be kept as small as possible under limited computational costs.
In our previous paper, assuming that the gravitational ringing is
dominated by the least-damped (fundamental) mode with the least
imaginary part of frequency, we constructed an efficient method for
tiling the template space. However, the dependence of the template
space metric on the initial phase of a wave was not taken into
account. This dependence arises because of an unavoidable mismatch
between the parameters of a signal waveform and those given discretely
in the template space.
In this paper, we properly take this dependence into account
and present an improved, efficient search method for gravitational
ringing of black holes.

%%%%%%%%%%%%%%%%%%%%%%%%%%%%%%%%%%%
%%%%%%%%%%%%%%%%%%%%%%%%%%%%%%%%%%%
\section{Introduction}
%%%%%%%%%%%%%%%%%%%%%%%%%%%%%%%%%%%
%%%%%%%%%%%%%%%%%%%%%%%%%%%%%%%%%%%

Thanks to the recent technological advance, 
there are many on-going projects of gravitational wave detection
in the world; the Laser Interferometric Gravitational 
Wave Observatory (LIGO)~\cite{LIGO}, VIRGO~\cite{VIRGO},
GEO-600~\cite{GEO}, ACIGA~\cite{ACIGA}, TAMA300~\cite{TAMA} 
and the Large-scale Cryogenic Gravitational wave Telescope (LCGT)~\cite{LCGT} 
which are ground-based laser interferometers, 
and EXPLORER~\cite{EXPLORER}, ALLEGRO~\cite{ALLEGRO}, NIOBE~\cite{NIOBE}, 
NAUTILUS~\cite{NAUTILUS} and AURIGA~\cite{AURIGA} which are bar detectors. 
Furthermore, there are some future space interferometer projects such 
as the Laser Interferometer Space Antenna (LISA)~\cite{LISA} and 
the DECi hertz Interferometer Gravitational wave
 Observatory (DECIGO)~\cite{DECIGO}. 
The detection of gravitational waves provides us with not only a direct 
experimental test of general relativity but also 
a new window to observe our Universe. 
To use them as a new tool of observation, it is necessary to 
derive theoretical waveforms. Once we know them, 
we may appeal to the matched filtering technique 
to extract source's information from gravitational wave signals. 
However, because the signals are expected to be 
very weak and the amount of data will be enormous 
for long-term continuous observations, it is essentially important 
to develop efficient data analysis methods. 

For these ground-based as well as future space-based interferometers, 
coalescence of compact object binaries is the most important source
of gravitational waves. The process of a binary coalescence can be
divided into three distinct phases. 
During the initial {\it inspiral} phase, the gravitational radiation 
reaction timescale is much longer than the orbital period. 
The gravitational waves from the inspiral phase carry the information of 
the masses, spins and so on, of the system.
After the inspiral phase, the binary becomes dynamically unstable
and starts to merge. 
This phase is called the {\it merger} phase. 
The gravitational waves from the merger phase give us 
the information about fully general relativistic dynamics of 
the system. Finally, if a black hole is formed
after the merger phase,
the system enters the {\it ringdown} phase where it gradually settles
down to a stationary Kerr black hole. During this process,
the black hole emits gravitational waves with frequencies and damping 
rates specific to its mass and spin.
Thus, the gravitational waves in the ringdown phase carry
the information of the mass and spin of the final black hole.

In this paper, we consider an effective search method for 
gravitational ringing of distorted spinning (Kerr) black holes. 
The ringdown waves are described by quasi-normal modes of 
a black hole.  The quasi-normal modes are complex frequency wave
solutions of the perturbed Einstein equations with purely 
outgoing-wave boundary condition at infinity and ingoing-wave at
horizon, with vanishing incoming-wave amplitude. 
A quasi-normal mode is characterized by the central frequency $f_c$, 
usually called the (quasi-)normal-mode frequency, 
and the quality factor $Q$ which is inversely proportional to 
the imaginary part of the complex frequency. 
Their properties were analyzed extensively by Leaver~\cite{Leaver}, 
and it is known that the least-damped (fundamental) mode belongs to the 
$\ell=m=2$ spin-$2$ spheroidal harmonic modes. 
we note that a ringdown signal decays exponentially. 
Therefore, unless the signal-to-noise ratio (SNR) is very large,
the signal will be soon buried in the noise after a few
oscillation cycles. So, it is essentially important to keep the loss
of SNR as small as possible when we construct a search method.

Here we consider a black hole characterized by its mass $M$ and 
the dimensionless spin parameter $a=J/M^2$ where $J$
is the spin angular momentum. The parameter $a$
takes a value in the range $[0,1)$, with $a=0$ corresponding to
 a Schwarzschild black hole 
and $a=1$ to an extreme Kerr black hole. 
It is known that the imaginary part of 
the $\ell=m=2$ least-damped mode 
is the smallest of all the quasi-normal modes, and 
results of black hole perturbation calculations as well as 
numerical relativity simulations strongly suggest that 
a ringdown wave is dominated by this 
$\ell=m=2$ least-damped mode unless $a$ 
is extremely close to unity. Hence we focus on this single mode. 
Then a ringdown waveform is expressed as 
\begin{eqnarray}
h(f_c,\,Q,\,t_0,\,\phi_0;\,t) = \cases{
e^{ - \frac {\pi \,f_c\,(t-t_0)}{Q}}\,\cos(2\,\pi \,f_c\,(t-t_0)-\phi_0) 
& for 
$t \geq t_0$ \,,
\cr 
0 & for $t < t_0$ \,,
\cr}
\label{eq:RDwave}
\end{eqnarray}
where we set the amplitude to unity for simplicity, 
and $t_0$ and $\phi_0$ are the initial time and phase of 
the ringdown wave, respectively. 
For the $\ell=m=2$ least-damped mode, analytical fitting formulas 
for the central frequency $f_c$ and quality factor $Q$
were given by Echeverria~\cite{ech} as 
\begin{eqnarray}
f_c &\simeq& 32{\rm kHz}\,[1-0.63\,(1-a)^{0.3}]
\left({M \over M_{\odot}}\right)^{-1} \,,
\label{eq:fcMa} 
\\
Q &\simeq& 2.0\,(1-a)^{-0.45} \,.
\label{eq:QMa}
\end{eqnarray}

There exists rich literature on search methods for ringdown waves.
Echeverria~\cite{ech} investigated 
the problem of extracting the black hole parameters 
from gravitational wave data in the case when SNR is large. 
Finn~\cite{finn} improved this situation by developing 
a maximum likelihood analysis method that can deal with any SNR. 
Flanagan and Hughes then considered the parameter extraction 
from the three stages of a binary coalescence, 
i.e., from inspiral, merger and ringdown phases, 
in their series of papers~\cite{FH}. 
For the ringdown phase, they discussed the relation between 
the energy spectrum of the radiation and SNR. 
Creighton~\cite{crei} reported the result of analyzing 
data of the Caltech 40m by matched filtering,
and emphasized the importance of coincidence event searches 
to discriminate spurious events from real events. 
But the search was limited to a single ringdown wave template. 
Recently, Arnaud et al.~\cite{arna} discussed a tiling method 
to cover the 2-dimensional template space $\{f_c,\,Q\}$. 
In our previous paper~\cite{NTTS}, we proposed a more efficient method
for tiling the template space. There, however, we ignored the dependence
of the metric of the template space $\{f_c,\,Q\}$ on the initial phase
$\phi_0$. This induces some small but non-negligible decrease in the 
match for a signal with certain ranges of $\phi_0$.
In this paper, we remove this shortcoming by properly taking
into account the initial phase dependence, and
develop a similar but substantially improved template spacing
which is much more reliable than the previous one.

Here we make a comment on an analysis using real interferometers' data. 
In this case, we have to deal with non-stationary, non-Gaussian noises,
and Tsunesada and Kanda~\cite{TsuneKanda} found that more
 fake events are observed than in the case of 
an inspiraling wave search, because the duration of a ringdown wave is
typically much shorter than that of an inspiral wave and
it can easily be affected by short bursts.
It will be necessary develop a way to remove such fake events 
without losing real ringdown signals, but we leave this issue
for future work. 

The paper is organized as follows. 
In Sec.~\ref{sec:TDA}, first, we introduce
 orthonormal template waveforms for ringdown waves. 
Second, assuming a white noise background, we consider matched filtering
in the 4-dimensional template space $\{t_0,\,\phi_0,\,f_c,\,Q\}$. 
We show that we can effectively reduce the template space to
2-dimensions spanned by $\{f_c,Q\}$, but the metric of this
reduced template space depends on $\phi_0$.
Then, by carefully taking account of the $\phi_0$ dependence of
the metric, we analytically develop an efficient and 
reliable tiling method for ringdown wave searches.
In Sec.~\ref{sec:TN}, by using a fitting curve for the TAMA noise spectrum 
during the Data Taking 8 (DT8) in 2003, 
we show that our template spacing developed for white noise 
is valid even in the case of colored noise. Finally,
Sec.~\ref{sec:Dis} is devoted to summary and discussion. 
In Appendix~\ref{app:TM}, we recapitulate the tiling method
we proposed in~\cite{NTTS}.
In Appendix~\ref{app:PEE}, we summarize the
parameter estimation errors for ringdown signals
by using the Fisher information matrix.

%%%%%%%%%%%%%%%%%%%%%%%%%%%%%%%%%%%
%%%%%%%%%%%%%%%%%%%%%%%%%%%%%%%%%%%
\section{Template space}\label{sec:TDA}
%%%%%%%%%%%%%%%%%%%%%%%%%%%%%%%%%%%
%%%%%%%%%%%%%%%%%%%%%%%%%%%%%%%%%%%

In this section, we present an improved version of the template
spacing we have developed in our previous paper~\cite{NTTS},
which can be used for matched filtering of the 
quasi-normal ringing waveforms.
Here, the detector noise is assumed to be white noise to make
it possible to deal with the problem analytically. 

%%%%%%%%%%%%%%%%%%%%%%%%%%%%%%%%%%%
\subsection{Maximization over the initial phase $\bm{\phi_0}$}
%%%%%%%%%%%%%%%%%%%%%%%%%%%%%%%%%%%

We have temporarily set the amplitude to unity for simplicity 
in Eq.~(\ref{eq:RDwave}). 
Note that the knowledge of the amplitude is not necessary
for the template spacing in matched filtering.
For $t \geq t_0$, 
the ringdown wave (\ref{eq:RDwave}) is divided into two parts. 
\begin{eqnarray}
h(f_c,\,Q,\,t_0,\,\phi_0;\,t) &=& h_c(f_c,\,Q,\,t_0;\,t)\cos \phi_0 
+ h_s(f_c,\,Q,\,t_0;\,t)\sin \phi_0 \,, 
\end{eqnarray}
where
\begin{eqnarray}
h_c(f_c,\,Q,\,t_0;\,t) &=& 
e^{ - \frac {\pi \,f_c\,(t-t_0)}{Q}}\,\cos(2\,\pi \,f_c\,(t-t_0)) \,,
\\
h_s(f_c,\,Q,\,t_0;\,t) &=& 
e^{ - \frac {\pi \,f_c\,(t-t_0)}{Q}}\,\sin(2\,\pi \,f_c\,(t-t_0)) \,.
\end{eqnarray}

Performing the Fourier transformation 
$\tilde{h}_{\sigma}(f)=\int_{-\infty}^{\infty}dt\,e^{2\pi ift}h_{\sigma}(t)$,
 where $\sigma=c$ or $s$,
we obtain the waveforms in the frequency domain as 
\begin{eqnarray}
\tilde{h}_c(f_c,\,Q,\,t_0;\,f) &=& {\displaystyle \frac 
{( f_c - 2\,i\,f\,Q)\,Q \,e^{2\,i\,\pi\,f\,t_0}}
{\pi(  2\,f_c\,Q - i\,f_c - 2\,f\,Q)\,(2\,f_c\,Q + i\,f_c 
+ 2\,f\,Q)}} \,, \label{eq:FHC} \\ 
\tilde{h}_s(f_c,\,Q,\,t_0;\,f) &=& {\displaystyle \frac 
{2\,f_c\,Q^2\,\,e^{2\,i\,\pi\,f\,t_0}}
{\pi(  2\,f_c\,Q - i\,f_c - 2\,f\,Q)\,(2\,f_c\,Q + i\,f_c 
+ 2\,f\,Q)}} \,. \label{eq:FHS} 
\end{eqnarray}
The waveform in the time domain is real, 
so the following relation is satisfied. 
\begin{eqnarray}
\tilde{h}^*(f)=\tilde{h}(-f) \,,
\end{eqnarray}
where the star (${~}^*$) denotes the complex conjugation. 

Here, we introduce the inner product as
\begin{eqnarray}
(a,\,b) = 2\,\int_{-f_{\rm max}}^{f_{\rm max}} df \,
{\tilde{a}(f) \tilde{b}^*(f) \over S_n(|f|)} \,,
\label{eq:inner}
\end{eqnarray}
where $S_n$ is defined as a one-sided power spectrum density, and
$f_{\rm max}$ is the maximum frequency we take into account in the analysis. 
In the actual data analysis, it is equal to or less than 
the half of the sampling frequency of data. 
In this section, the detector noise is assumed 
to be white noise $S_n(|f|)=1$. We will use the fitting curve 
for the TAMA DT8 noise spectrum as a colored noise in the next section. 

In the matched filtering, we calculate the inner product 
between the template $h$ and a signal $x$ defined by $(x,h)$. 
First, we normalize the templates $h_c$ and $h_s$. 
We define the normalization constants as 
\begin{eqnarray}
N_c(f_c,\,Q,\,t_0) &=& 
({h}_c(f_c,\,Q,\,t_0),{h}_c(f_c,\,Q,\,t_0))
\nonumber \\ 
&=& {\displaystyle \frac 
{(2\,Q^{2} + 1)\,Q}
{\pi \,(4\,Q^{2} + 1)\,f_c}} \,, \\
N_s(f_c,\,Q,\,t_0) &=& 
({h}_s(f_c,\,Q,\,t_0),{h}_s(f_c,\,Q,\,t_0))
\nonumber \\ 
&=& {\displaystyle \frac 
{2\,Q^3}
{\pi \,(4\,Q^{2} + 1)\,f_c}} \,,
\end{eqnarray}
when we have set $f_{\rm max}=\infty$. 
It is noted that the normalization constant $N_{\sigma}$ 
does not depend on the initial time. 
The normalized templates $\hat{h}_{\sigma}(f_c,\,Q,\,t_0)$ are
given by 
\begin{eqnarray}
\tilde{\hat{h}}_{\sigma}(f_c,\,Q,\,t_0;\,f) 
= {1 \over \sqrt{N_{\sigma}(f_c,\,Q,\,t_0)}}
 \tilde{h}_{\sigma}(f_c,\,Q,\,t_0;\,f) \,.
\end{eqnarray}
We note that the two parts, $h_c$ and $h_s$, are not orthogonal. 
Their inner product is obtained as 
\begin{eqnarray}
(\hat{h}_c(f_c,\,Q,\,t_0),\hat{h}_s(f_c,\,Q,\,t_0)) 
&=& {1 \over \sqrt{2\,(2\,Q^2+1)}} 
\nonumber \\ &=:& c(f_c,\,Q,\,t_0) \,.
\end{eqnarray}
Then the normalized ringdown template for the waveform~(\ref{eq:RDwave})
is given in the frequency domain as
\begin{eqnarray}
\tilde{\hat{h}}(\phi_0;\,f_c,\,Q,\,t_0;\,f) 
&=& {1 \over \sqrt{N(\phi_0;\,f_c,\,Q,\,t_0)}}
 \tilde{h}(f_c,\,Q,\,t_0,\,\phi_0;\,f) \,,
\end{eqnarray}
where
\begin{eqnarray}
\tilde{h}(f_c,\,Q,\,t_0,\,\phi_0;\,f) 
&=& {\displaystyle \frac {( \cos \phi_0\,f_c - 2\,i\,\cos \phi_0\,f\,Q 
+ 2\,f_c\,Q\,\sin \phi_0)\,Q\,e^{ 2\,i\,\pi \,f\,t_0 }}
{\pi\, ( 2\,f_c\,Q - i\,f_c - 2\,f\,Q)\,(2\,f_c\,Q + i\,f_c + 2\,f\,Q)}} \,,
\nonumber \\ 
N(\phi_0;\,f_c,\,Q,\,t_0) &=& 
({h}(f_c,\,Q,\,t_0,\,\phi_0),{h}(f_c,\,Q,\,t_0,\,\phi_0))
\nonumber \\ 
&=& 
N_c(f_c,\,Q,\,t_0) \cos^2 \phi_0 + N_s(f_c,\,Q,\,t_0) \sin^2 \phi_0 
\nonumber \\ &&
+ 2\,c(f_c,\,Q,\,t_0)\sqrt{N_c(f_c,\,Q,\,t_0)N_s(f_c,\,Q,\,t_0)}
\cos \phi_0 \sin \phi_0 
\,.
\label{eq:NGwave}
\end{eqnarray}

In an actual data analysis, instead of the normalized templates
 $\hat{h}_{c}$ and $\hat{h}_{s}$, 
it is more convenient to prepare a set of
orthonormalized waveforms $h_1$ and $h_2$ as templates. 
They can be obtained by Schmidt's orthonormalization. 
First, we choose 
\begin{eqnarray}
{h}_1(f_c,\,Q,\,t_0) = \hat{h}_c(f_c,\,Q,\,t_0) \,.
\label{eq:h1}
\end{eqnarray}
Then the normalized waveform $h_2$ orthogonal to
 $h_1$ is obtained as 
\begin{eqnarray}
{h}_2(f_c,\,Q,\,t_0) = 
{\hat{h}_s(f_c,\,Q,\,t_0) 
- c(f_c,\,Q,\,t_0) \hat{h}_c(f_c,\,Q,\,t_0) 
\over \sqrt{1-c(f_c,\,Q,\,t_0)^2}} \,.
\label{eq:h2}
\end{eqnarray}
Using the above orthonormal templates, the SNR maximized over the 
initial phase $\phi_0$ is given as 
\begin{eqnarray}
\rho(x;f_c,\,Q,\,t_0) 
=\mathop{\rm Max}\limits_{\phi_0}
\left\{\sqrt{(x,\hat{h}(f_c,\,Q,\,t_0,\,\phi_0))^2}\right\}
= \sqrt{(x,\,{h}_1(f_c,\,Q,\,t_0))^2\vphantom{\hat{h}}
+(x,\,{h}_2(f_c,\,Q,\,t_0))^2} \,,
\label{eq:rho}
\end{eqnarray}
where $x$ represents a signal.
We note that there is basically nothing wrong with
using the non-orthogonalized templates $\hat{h}_{\sigma}$ 
in the matched filtering. In the case we use the non-orthogonal templates, 
the maximization of $(x,\hat{h})$ over the phase $\phi_0$ has been 
discussed by Mohanty~\cite{Moh}, and the result is the same if
we replace $h_1$ and $h_2$ in Eq.~(\ref{eq:rho}) by those given in
Eqs.~(\ref{eq:h1}) and (\ref{eq:h2}), respectively.

Before closing this subsection, let us describe how to determine
the initial phase $\phi_0$ from the filtered output.
For simplicity, we take the normalized wave $\hat{h}$ as a signal,
because the overall amplitude is irrelevant for the present discussion.
We further assume that there is a perfect match of the parameters
$\{f_c,\,Q,\,t_0\}$ between the signal and a template, and
the SNR is maximized, that is, $\rho(\hat{h};f_c,\,Q,\,t_0)=1$.

We consider the quantities,
\begin{eqnarray}
\rho_1 &=& (\hat{h},\,{h}_1) \,, \nonumber \\
\rho_2 &=& (\hat{h},\,{h}_2) \,. 
\end{eqnarray}
When the signal perfectly matches with a template,
these are formally expressed as
\begin{eqnarray}
\rho_1 &=&  {1 \over \sqrt{N(\phi_0;\,f_c,\,Q,\,t_0)}} 
\left(\sqrt{N_c(f_c,\,Q,\,t_0)} \cos \phi_0 
+ c(f_c,\,Q,\,t_0) \sqrt{N_s(f_c,\,Q,\,t_0)} \sin \phi_0 \right) \\
\nonumber \\ 
\rho_2 &=& {1 \over \sqrt{N(\phi_0;\,f_c,\,Q,\,t_0)}} 
\sqrt{(1-c(f_c,\,Q,\,t_0)^2) N_s(f_c,\,Q,\,t_0)} \sin \phi_0 \,.
\end{eqnarray}
Taking the ratio of $\rho_1$ and $\rho_2$, we find
\begin{eqnarray}
{\rho_1 \over \rho_2} = 
\sqrt{{N_c(f_c,\,Q,\,t_0) \over \left(1-c(f_c,\,Q,\,t_0)^2\right)
 N_s(f_c,\,Q,\,t_0)}} \cot \phi_0 
+\sqrt{{c(f_c,\,Q,\,t_0) \over 1-c(f_c,\,Q,\,t_0)^2}} \,.
\end{eqnarray}
Thus, the initial phase of a ringdown wave signal is determined as 
\begin{eqnarray}
\phi_0 = \cot^{-1} 
\left[ 
\sqrt{{\left(1-c(f_c,\,Q,\,t_0)^2\right)
 N_s(f_c,\,Q,\,t_0) \over N_c(f_c,\,Q,\,t_0)}}
{\rho_1 \over \rho_2} 
- \sqrt{{c(f_c,\,Q,\,t_0) N_s(f_c,\,Q,\,t_0) \over N_c(f_c,\,Q,\,t_0)}}
\right] \,. 
\end{eqnarray} 
It is noted that $\phi_0 \sim \cot^{-1}(\rho_1/\rho_2)$ for large $Q$. 

%%%%%%%%%%%%%%%%%%%%%%%%%%%%%%%%%%%
\subsection{3-dimensional distance function}
%%%%%%%%%%%%%%%%%%%%%%%%%%%%%%%%%%%

In the previous subsection, we obtained the expression for the
SNR maximized over the initial phase $\phi_0$, Eq.~(\ref{eq:rho}).
Here, we consider the remaining 3 parameters $\{f_c,\,Q,\,t_0\}$, 
and derive the metric in the 3-dimensional template space that
describes the degree of mismatch between a signal and a template.
To this end, we point out an important fact that was overlooked 
in~\cite{NTTS}. Given a signal with certain values of
the parameters, the SNR defined by Eq.~(\ref{eq:rho}) will
be independent of $\phi_0$ for a template that matches exactly 
with the signal. However, it will not be so if there is no exact
matching between the signal and a template.
In other words, if there is a mismatch between the parameters
of a signal and those of a template, the resulting SNR will
depend on the initial phase $\phi_0$.

Let us define the match $C(\phi_0;\,df_c,\,dQ,\,dt_0)$ 
between the template with the parameters $(f_c,\,Q,\,t_0,\,\phi_0)$ and
the normalized signal having slightly different values of the
parameters $(f_c+d f_c,\,Q+dQ,\,t_0+dt_0,\,\phi_0)$.
Note that, since we have already maximized the SNR over
the initial phase $\phi_0$, it is unnecessary to consider the
difference $d\phi_0$ for the match.
The match $C(\phi_0;\,df_c,\,dQ,\,dt_0)$ is defined by 
\begin{eqnarray}
C(\phi_0;\,d f_c,\,dQ,\,dt_0) &:=& 
\rho\left(\hat{h}(f_c+df_c,\,Q+dQ,\,t_0+dt_0,\,\phi_0);f_c,\,Q,\,t_0\right)
\nonumber \\ &=& 
\left[ \sum_{A=1}^2
\frac{\bigl({h}(f_c+d f_c,\,Q+dQ,\,t_0+dt_0,\,\phi_0),
{h}_A(f_c,\,Q,\,t_0)\bigr)^2}
{\bigl({h}(f_c+d f_c,\,Q+dQ,\,t_0+dt_0,\,\phi_0),\,
{h}(f_c+d f_c,\,Q+dQ,\,t_0+dt_0,\,\phi_0)\bigr)} \right]^{1/2}
\nonumber \\ 
&=& 
1 - {1 \over 2 ({h}(f_c,\,Q,\,t_0,\,\phi_0),\,{h}(f_c,\,Q,\,t_0,\,\phi_0))} 
\biggl[({h}_{,i}(f_c,\,Q,\,t_0,\,\phi_0),\,{h}_{,j}(f_c,\,Q,\,t_0,\,\phi_0))
\nonumber \\ && 
-\sum_{A=1}^{2} ({h}_{,i}(f_c,\,Q,\,t_0,\,\phi_0),\,
{h}_A(f_c,\,Q,\,t_0)) 
({h}_{,j}(f_c,\,Q,\,t_0,\,\phi_0),\,{h}_A(f_c,\,Q,\,t_0)) 
\biggr] dx^idx^j 
\nonumber \\ && + O(dx^3) 
\nonumber \\ &=& 
1-ds^2_{(3)} + O(dx^3)\,,
\label{eq:match}
\end{eqnarray}
where $\{x^i\}=\{t_0,\,f_c,\,Q\}$ ($i=1,2,3$),
the comma ($,$) denotes the partial differentiation,
and we have introduced the 3-dimensional distance function 
in the template space by $ds^2_{(3)}=1-C$. 
The smaller the match $C$ is, the larger the distance is 
between the two signals in the template space.

If we introduce the metric~\cite{owen} as
\begin{eqnarray}
ds^2_{(3)}=g_{ij}^{(3)}dx^idx^j \,,
\end{eqnarray}
the components of the metric are explicitly written as 
\begin{eqnarray}
g_{t_0 t_0}^{(3)} &=& 
\frac{4\,\pi\,{f_c}\,{f_{\rm max}} \,\left( 4\,Q^2 + 1 \right) \,
{\cos^2 (\phi_0)}}{Q\,
\left( 4\,Q^2 + 2\,Q\,\sin (2\,\phi_0) + 1 + \cos (2\,\phi_0)  \right) }
\,,
\nonumber \\ 
g_{t_0 f_c}^{(3)} &=& g_{f_c t_0}^{(3)} 
\nonumber \\ &=& 
\frac{\pi \,\left( 4\,Q^2 + 1 \right) \,\sin (2\,\phi_0)}
  {2\,\left( 4\,Q^2 + 2\,Q\,\sin (2\,\phi_0) + 1 + \cos (2\,\phi_0)  \right) }
\,,
\nonumber \\ 
g_{t_0 Q}^{(3)} &=& g_{Q t_0}^{(3)} 
\nonumber \\ &=& 
\frac{\pi \,{f_c}\,\cos (\phi_0)\,
    \left( 2\,Q\,\cos (\phi_0) - \sin (\phi_0) \right) }{Q\,
    \left( 4\,Q^2 + 2\,Q\,\sin (2\,\phi_0) + 1 + \cos (2\,\phi_0)  \right) }
\,,
\nonumber \\ 
g_{f_c f_c}^{(3)} &=& 
\frac{Q^2\,\left( 4\,Q^2 - 2\,Q\,\sin (2\,\phi_0) + 1 - \cos (2\,\phi_0)  \right)}
{2\,{f_c}^2\,\left( 4\,Q^2 + 2\,Q\,\sin (2\,\phi_0) + 1 + \cos (2\,\phi_0)  \right) }
\,,
\nonumber \\ 
g_{f_c Q}^{(3)} &=& g_{Q f_c}^{(3)} 
\nonumber \\ &=& 
-\frac{Q\,{\left( 2\,Q\,\cos (\phi_0) - \sin (\phi_0) \right)}^2}
{{f_c}\,\left( 4\,Q^2 + 1 \right) \,
      \left( 4\,Q^2 + 2\,Q\,\sin (2\,\phi_0) + 1 + \cos (2\,\phi_0)  \right) }
\,, 
\nonumber \\ 
g_{QQ}^{(3)} &=& 
\frac{{\left( 4\,Q^2 + 1 \right) }^2 + 
    \left( 12\,Q^2 - 1 \right) \,\cos (2\,\phi_0) + 
    2\,Q\,\left( 4\,Q^2 - 3 \right) \,\sin (2\,\phi_0)}{2\,
    {\left( 4\,Q^2 + 1 \right) }^2\,
    \left( 4\,Q^2 + 2\,Q\,\sin (2\,\phi_0) + 1 + \cos (2\,\phi_0)  \right) }
\,,
\end{eqnarray}
where we have assumed $f_{\rm max}\gg f_c$ and retained
only the leading terms in $f_{\rm max}/f_c$. Note that
$f_{\rm max}$ appears only in the component $g^{(3)}_{t_0t_0}$
in this limit.
In the next subsection, we shall argue that the dependence on
$f_{\rm max}$ will disappear from the final result if
 the maximization over
$t_0$ is done with accuracy $dt_0\lesssim 1/f_{\rm max}$.

The inequality $C(\phi_0;\,d f_c,\,dQ,\,dt_0) \leq 1$ means that 
there will be a loss of SNR unless the actual
parameters of a gravitational wave signal fall exactly
onto one of the templates. 
In order to keep the detection efficiency high enough,
we have to choose the template spacing so that the
maximum value of the distance function is kept
as small as possible.

%%%%%%%%%%%%%%%%%%%%%%%%%%%%%%%%%%%
\subsection{Projection to 2-dimensional template space}\label{subsec:P2}
%%%%%%%%%%%%%%%%%%%%%%%%%%%%%%%%%%%

The 3-dimensional metric on the template space $\{t_0,\,f_c,\,Q\}$ has
a simple $f_{\rm max}$ dependence in the large $f_{\rm max}$ limit,
namely, $g^{(3)}_{t_0t_0}\propto f_{\rm max}$ and the other components
are independent of $f_{\rm max}$. This fact allows us to reduce the
problem of template spacing to that in the 2-dimensional space
$\{f_c,\,Q\}$.

To reduce the 3-dimensional template space to 2-dimensions, let us
note the distance function $ds_{(3)}^2$ may be rewritten as~\cite{owen}
\begin{eqnarray}
ds_{(3)}^2
=g^{(3)}_{t_0t_0}
\left(dt_0+\frac{g^{(3)}_{t_0I}}{g^{(3)}_{t_0t_0}}dx^I\right)^2
+g_{IJ}dx^Idx^J\,,
\end{eqnarray}
where $\{x^I\}=\{f_c,\,Q\}$ and
\begin{eqnarray}
g_{IJ} = g^{(3)}_{IJ} 
- \frac{g^{(3)}_{t_0I}g^{(3)}_{t_0J}}{g^{(3)}_{t_0 t_0}} \,.
\label{eq:3to2}
\end{eqnarray}
We see that we can minimize the distance function (maximize the match)
with respect to $t_0$ by filtering a data stream with respect to $t_0$ 
with accuracy $dt_0=O(1/f_{\rm max})$, because we then have
\begin{eqnarray}
g^{(3)}_{t_0t_0}
\left(dt_0+\frac{g^{(3)}_{t_0I}}{g^{(3)}_{t_0t_0}}dx^I\right)^2
=O(f_c/f_{\rm max})\to 0\,,
\end{eqnarray}
in the limit $f_{\rm max}/f_c\gg 1$. Thus, provided the matched
filtering is done with this accuracy, which is normally practiced
when we sweep over a data stream in time sequence,
we may focus on the 2-dimensional space $\{f_c,\,Q\}$ for the
problem of template spacing.
Note that in the limit $f_{\rm max}/f_c\to\infty$,
the reduced 2-dimensional distance function is given by 
\begin{eqnarray}
ds^2_{(2)} =g_{IJ}dx^{I}dx^{J}=g^{(3)}_{IJ}dx^Idx^J \,.
\label{eq:ds2}
\end{eqnarray}
Thus the $f_{\rm max}$ dependence disappears completely.

Before proceeding further, we note that the metric $g_{IJ}$ has
a simple central frequency dependence which may be removed
by the coordinate transformation,
\begin{eqnarray}
dF &=& \frac{d f_c}{f_c} \,;
\nonumber \\ 
f_c &\to& F=\ln (f_c/f_0) \,,
\nonumber \\ 
F &\to& f_c= f_0\,e^{F} \,,
\label{eq:defdF}
\end{eqnarray}
where $f_0$ is some fiducial frequency. 
This transformation gives
\begin{eqnarray}
ds^2_{(2)} &=& g_{FF} \,dF^{2} + 2\,g_{FQ} \,dF\,dQ + g_{QQ} \,dQ^{2} 
\,;
\nonumber \\ 
g_{FF} &=& f_c^2\, g_{f_c f_c} \,, \quad g_{FQ} = f_c\, g_{f_c Q} \,. 
\label{eq:ds2NN}
\end{eqnarray}

\begin{figure}[ht]
\begin{center}
\epsfxsize=8cm
\hspace{-1cm}\epsfbox{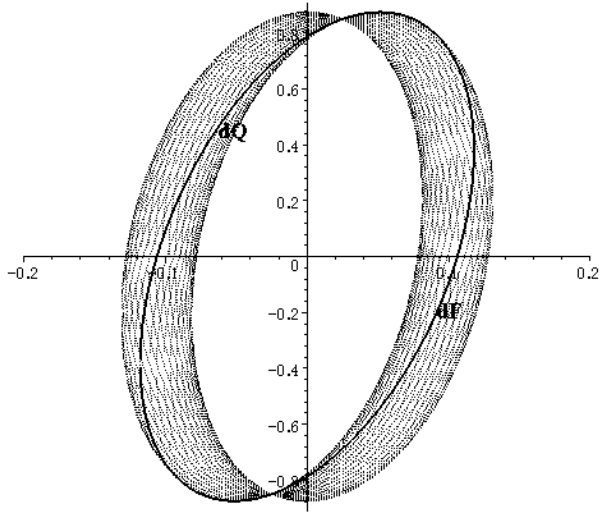}
\caption{Contours of the maximum distance $ds^2_{\rm max}=0.02$
for various initial phases. The quality factor is set to $Q=2$. 
The thick curve shows the case of $\phi_0=0$, the interior of
which corresponds to the region covered by one template in~\cite{NTTS}. }
\label{fig:rinF1}
\end{center}
\end{figure}

Now, to find an appropriate tiling of the 2-dimensional
template space, we consider contours of a fixed maximum distance
$ds^2_{(2)}=ds^2_{\rm max}$. Note that the metric has the
initial phase dependence. So, the contour varies as
$\phi_0$ is varied (over the range $0 \leq \phi_0 \leq \pi$).
The contours for $ds^2_{\rm max}=0.02$ are plotted in
Fig.~\ref{fig:rinF1} in the case of the quality factor $Q=2$.
The thick curve shows the contour with $\phi_0=0$ which we considered 
in~\cite{NTTS}. 
We see that if we use the previous template spacing given there, 
the minimum match is not guaranteed for other values of the initial
phase $\phi_0$.
It is therefore necessary to remedy this drawback when tiling the
template space. 

%%%%%%%%%%%%%%%%%%%%%%%%%%%%%%%%%%%
\subsection{2-dimensional template spacing}
%%%%%%%%%%%%%%%%%%%%%%%%%%%%%%%%%%%

We now consider the template spacing that guarantees the minimum match
for any value of $\phi_0$. As seen from Fig.~\ref{fig:rinF1}, 
a conservative method is to take the interior of the inner
envelope curve of the contours to be a region covered by each template.
If we adopt such a template spacing based on the exact inner envelope curve,
the computational cost of assigning grid points in the template space 
becomes high.
We therefore choose to allow some redundancy, and construct an ellipse
that can cover the interior of the inner envelope curve as much as
possible. This also enables us to use the efficient method of template
space tiling developed in~\cite{NTTS}.

\begin{figure}[ht]
\begin{center}
\epsfxsize=8cm
\hspace{-1cm}\epsfbox{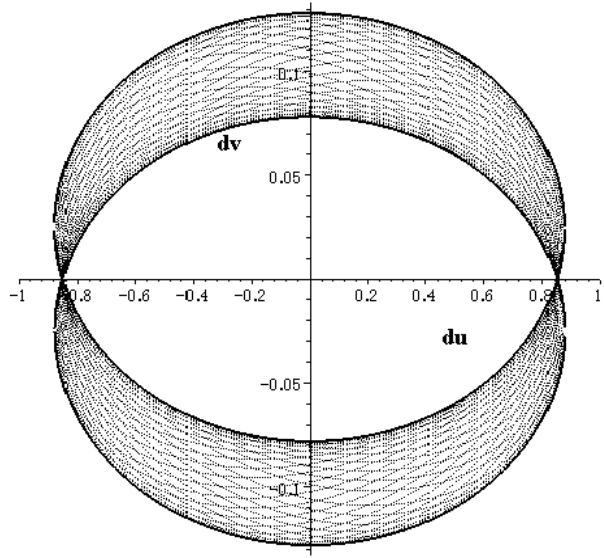}
\caption{Contours of the fixed maximum distance $ds^2_{\rm max}=0.02$
for $Q=2$ on the $(u,v)$-plane with the metric~(\ref{eq:ds2uv}).
}
\label{fig:rinF2}
\end{center}
\end{figure}

Let us determine the inner envelope curve. It is convenient to
perform a coordinate transformation that makes the envelope curve 
symmetric with respect the axes of new coordinates.
To do so, we first make a rotation in the $(F,Q)$-plane to bring
the two points of intersection of all the contours on to the axes.
This is achieved by the coordinate transformation
\begin{eqnarray}
du &=& \cos \theta \,dF + \sin \theta \,dQ \,,
\nonumber \\ 
dv &=& - \sin \theta \,dF + \cos \theta \,dQ \,,
\end{eqnarray}
 where 
\begin{eqnarray}
\cos \theta = {1 \over \sqrt{16\,Q^6+8\,Q^4+Q^2+1}} \,, 
\nonumber \\ 
\sin \theta = {Q\,(4\,Q^2+1) \over \sqrt{16\,Q^6+8\,Q^4+Q^2+1}} \,. 
\end{eqnarray}
The metric components in the new coordinates are given by
\begin{eqnarray}
ds^2_{(2)} &=& g_{uu} \,du^{2} + 2\,g_{uv} \,du\,dv + g_{vv} \,dv^{2} 
\,;
\nonumber \\ 
g_{uu} &=& 
\cos^2 \theta \,g_{FF} + 2\,\cos \theta \,\sin \theta \,g_{FQ}
+\sin^2 \theta \,g_{QQ} \,,
\nonumber \\ 
g_{uv} &=& 
-\cos \theta \,\sin \theta \,g_{FF} 
+ (\cos^2 \theta - \sin^2 \theta) \,g_{FQ}
+\cos \theta \,\sin \theta \,g_{QQ} \,,
\nonumber \\ 
g_{vv} &=& 
\sin^2 \theta \,g_{FF} - 2\,\cos \theta \,\sin \theta \,g_{FQ}
+\cos^2 \theta \,g_{QQ} \,.
\label{eq:ds2uv}
\end{eqnarray}
Using the large $Q$ expansion valid for $Q \geq 2$, the new coordinates
$u$ and $v$ as functions of $Q$, to $O(1/Q^8)$ inclusive,
are given by
\begin{eqnarray}
u &=& \left[{1 \over 4\,Q^3}-{1 \over 16\,Q^5}
+{1 \over 64\,Q^7}\right]\,F
+\left[Q+{1 \over 160\,Q^5}-{1 \over 448\,Q^7}\right] \,,
\nonumber \\ 
v &=& -\left[1-{1 \over 32\,Q^6}+{1 \over 64\,Q^8}\right]\,F
+\left[-{1 \over 8\,Q^2}+{1 \over 64\,Q^4}
-{1 \over 384\,Q^6}+{3 \over 2048\,Q^8}\right] \,.
\end{eqnarray}
The inverse transformation is given by 
\begin{eqnarray}
F &=& \left[-1+{1 \over 32\,u^6}-{1 \over 64\,u^8}\right]\,v 
+\left[-{1 \over 8\,u^2}+{1 \over 64\,u^4}
-{1 \over 384\,u^6}+{39 \over 10240\,u^8}\right]
\,,
\nonumber\\
Q &=& -{3 \over 16\,u^7}\,v^2
+\left[{1 \over 4\,u^3}-{1 \over 16\,u^5}+{1 \over 64\,u^7}\right]\,v 
+\left[u+{1 \over 40\,u^5}-{17 \over 1792\,u^7}\right]\,.
\label{eq:invtrans}
\end{eqnarray}
The contours of the maximum distance $ds_{\rm max}^2=0.02$
for $Q=2$ on the $(u,v)$-plane are shown in Fig.~\ref{fig:rinF2}.

Next, we derive the envelope curves.
An envelope curve is the solution of the following two equations. 
\begin{eqnarray}
R(du,\,dv;\,\phi_0) &=& 0 \,,
\nonumber \\ 
{\partial \over \partial \phi_0} \,R(du,\,dv;\,\phi_0) &=& 0 \,,
\label{eq:houraku}
\end{eqnarray}
where the function $R$ is given in our situation by 
\begin{eqnarray}
R(du,\,dv;\,\phi_0)
= g_{uu} \,du^{2} + 2\,g_{uv} \,du\,dv + g_{vv} \,dv^{2} - ds^2_{\rm max} 
\,.
\end{eqnarray}
The envelope curve is obtained by eliminating the 
initial phase $\phi_0$ from Eqs.~(\ref{eq:houraku}).
We find
\begin{eqnarray}
\left[du+{1 \over Q\,(4\,Q^2+1)}dv\right]^2
&=&\frac{16\,Q^6+8\,Q^4+Q^2+1}{4\,Q^{5} (4\,Q^2+1)^2}
\nonumber \\ 
&& \times 
\biggl[2\,Q\,(16\,Q^4+8\,Q^2+1)\,ds^2_{\rm max} 
-Q\,(16\,Q^6+8\,Q^4+Q^2+1 )\,dv^2 
\nonumber \\ && \qquad 
\pm (4\,Q^2+1)\,\sqrt{2\,(16\,Q^6+8\,Q^4+Q^2+1)}\,ds_{\rm max}\,|dv|
\biggr]
\,,
%du &=& - {1 \over Q\,(4\,Q^2+1)}dv
%\pm { \sqrt{16\,Q^6+8\,Q^4+Q^2+1} \over 2\,Q^{5/2} (4\,Q^2+1)}
%\nonumber \\ && \times 
%\biggl[2\,Q\,(16\,Q^4+8\,Q^2+1)\,ds^2_{\rm max} 
%-Q\,(16\,Q^6+8\,Q^4+Q^2+1 )\,dv^2 
%\nonumber \\ && \qquad 
%\pm (4\,Q^2+1)\,\sqrt{2\,(16\,Q^6+8\,Q^4+Q^2+1)}\,ds_{\rm max}\,dv
%\biggr]^{1/2} 
%\,.
\label{eq:midhou}
\end{eqnarray}
where $Q$ is a function of $u$ and $v$ as given in Eq.~(\ref{eq:invtrans}).
The outer and inner envelope curves are given by 
the plus and minus signs, respectively, in front of the square root
in the above solution.
Since our interest is in the inner envelope curve, we focus
on the minus sign solution.

\begin{figure}[ht]
\begin{center}
\epsfxsize=8cm
\hspace{-1cm}\epsfbox{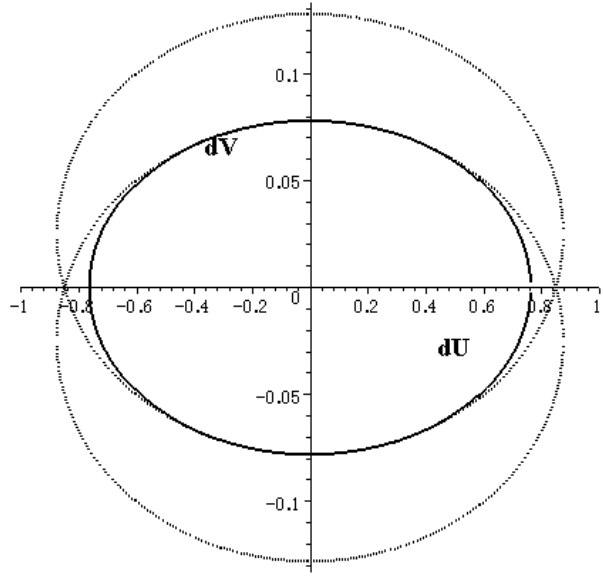}
\caption{The dotted curves show the inner and outer envelope curves
for the maximum distance $ds^2_{\rm max}=0.02$.
The thick curve is the ellipse given by Eq.~(\ref{eq:RELds2}).
 for all initial phases. 
Here we set $Q=2$.}
\label{fig:rinF3}
\end{center}
\end{figure}

As we can see from Eq.~(\ref{eq:midhou}),
the envelope curve is still asymmetric with respect to both axes.
To make it symmetric, we make a further coordinate transformation, 
\begin{eqnarray}
dV &=& dv \,, 
\nonumber \\ 
dU &=& du + {1 \over Q\,(4\,Q^2+1)}dv \,, 
\label{eq:toUV}
\end{eqnarray}
which gives the inner envelope curve in the form,
\begin{eqnarray}
dU^2&=&\frac{16\,Q^6+8\,Q^4+Q^2+1}{4\,Q^{5} (4\,Q^2+1)^2}
\nonumber\\
&&\times
\biggl[2\,Q\,(16\,Q^4+8\,Q^2+1)\,ds^2_{\rm max} 
-Q\,(16\,Q^6+8\,Q^4+Q^2+1 )\,dV^2 
\nonumber \\ && 
\qquad
-(4\,Q^2+1)\,\sqrt{2\,(16\,Q^6+8\,Q^4+Q^2+1)}\,ds_{\rm max}|dV|
\biggr]
\,.
\label{eq:lashou}
\end{eqnarray}
The equations (\ref{eq:toUV}) are integrated to give
\begin{eqnarray}
V &=& v
\nonumber \\ 
&=& -\left[1-{1 \over 32\,Q^6}+{1 \over 64\,Q^8}\right]\,F 
+\left[-{1 \over 8\,Q^2}+{1 \over 64\,Q^4}
-{1 \over 384\,Q^6}+{3 \over 2048\,Q^8}\right] 
\,, 
\nonumber \\ 
U &=& Q-{1 \over 160\,Q^5}+{1 \over 448\,Q^7} 
\,.
\end{eqnarray}
The inverse transformation is given by 
\begin{eqnarray}
F &=& 
\left[-1-{1 \over 32\,U^6}+{1 \over 64\,U^8}\right]\,V 
+\left[-{1 \over 8\,U^2}+{1 \over 64\,U^4}-{1 \over 384\,U^6}
-{9 \over 10240\,U^8}\right]
\,, 
\nonumber \\ 
Q &=& U+{1 \over 160\,U^5}-{1 \over 448\,U^7} 
\,. 
\end{eqnarray}
It is noted that the quality factor $Q$ depends on only $U$.
The inner and outer envelope curves are shown by the dotted curves in 
Fig.~\ref{fig:rinF3}.

Finally, to use the efficient tiling method developed in~\cite{NTTS}, 
we look for an ellipse inside the inner envelope curve that
can cover the region as much as possible.
A simplest choice is to construct the ellipse 
tangent to the envelope curve at $dU=0$ with the same curvature.
This gives
\begin{eqnarray}
ds^2_{\rm max} &=& 
{2\,Q^4 (\sqrt{4\,Q^2+1}+1)
\over \sqrt{4\,Q^2+1}\,(16\,Q^6+8\,Q^4+Q^2+1)} dU^2
\nonumber \\ && 
+ {Q^2\,(16\,Q^6+8\,Q^4+Q^2+1) \over 
(4\,Q^2+1)^2 (2\,Q^2+1 -\sqrt{4\,Q^2+1})}dV^2
\nonumber \\ 
&=:& g_{UU}\,dU^2 + g_{VV}\,dV^2
\,.
\label{eq:RELds2}
\end{eqnarray}
This inner ellipse is shown by the thick curve in Fig.~\ref{fig:rinF3}. 
It is noted that the ratio between the area covered by
the inner ellipse and that by the inner envelope curve 
is $95.0\%$ for $Q=2$ and $99.4\%$ for $Q=20$, 
and this ratio is independent of the maximum 
distance $ds_{\rm max}$ and the central frequency $f_c$.

%%%%%%%%%%%%%%%%%%%%%%%%%%%%%%%%%%%
\subsection{Conformally flat metric}\label{subsec:CFM}
%%%%%%%%%%%%%%%%%%%%%%%%%%%%%%%%%%%

Once we have an ellipse that ensures the minimum match we require,
it is straightforward to apply our tiling method developed in~\cite{NTTS}.
To do so, we perform a coordinate transformation that transforms
the 2-dimensional metric~(\ref{eq:RELds2}) to an explicitly conformally flat
form. Namely, by the transformation,
\begin{eqnarray}
dX &=& -dV \,,
\nonumber \\ 
dY &=& - \sqrt{{g_{UU} \over g_{VV}}} \,dU \,,
\end{eqnarray}
we obtain
\begin{eqnarray}
ds^2_{\rm max} =\Omega(Y) \left(dX^2+dY^2\right) \,,
\label{eq:ds2coN}
\end{eqnarray}
where the conformal factor is given by 
\begin{eqnarray}
\Omega(Y)=g_{VV}\bigl(Q(Y)\bigr) \,.
\label{eq:conformal}
\end{eqnarray}
Up to $O(1/Q^8)$ inclusive, the coordinate transformation is
explicitly given as 
\begin{eqnarray}
X &=& \left[1-{1 \over 32\,Q^6}+{1 \over 64\,Q^8}\right]\,F 
+\left[{1 \over 8\,Q^2}-{1 \over 64\,Q^4}
+{1 \over 384\,Q^6}-{3 \over 2048\,Q^8} \right] 
\,, 
\nonumber \\ 
Y &=& 
{1 \over 2\,Q}-{1 \over 16\,Q^2}
-{3 \over 64\,Q^3}+{11 \over 1024\,Q^4}
+{31 \over 4096\,Q^5}-{69 \over 32768\,Q^6}
\nonumber \\ && 
-{3357 \over 917504\,Q^7}+{3891 \over 4194304\,Q^8}
+{248531 \over 150994944\,Q^9} \,.
\end{eqnarray}
To the same accuracy, the inverse transformation becomes 
\begin{eqnarray}
F &=& \left[1+2\,Y^6+3\,Y^7+{31 \over 8}\,Y^8 \right] \,X 
\nonumber \\ &&
+\biggl[
-{1 \over 2}\,Y^2-{1 \over 4}\,Y^3
-{9 \over 32}\,Y^4-{1 \over 4}\,Y^5
-{25 \over 96}\,Y^6-{17 \over 64}\,Y^7
-{9447 \over 7168}\,Y^8 \biggr] 
\,,
\nonumber \\ 
Q &=& {1 \over 2\,Y}-{1 \over 8}
-{7 \over 32} \,Y-{9 \over 128}\,Y^2
-{31 \over 512}\,Y^3-{97 \over 2048}\,Y^4-{10649 \over 57344}\,Y^5
\nonumber \\ &&
-{7457 \over 32768}\,Y^6
-{373751 \over 1179648}\,Y^7 \,.
\end{eqnarray}
The conformal factor $\Omega(Y)$ is given by 
\begin{eqnarray}
\Omega(Y) &=& {1 \over 8\,Y^2}+{3 \over 16\,Y}
+{11 \over 128}+{1 \over 128}\,Y
-{3 \over 2048}\,Y^2+{1 \over 4096}\,Y^3+{98297 \over 229376}\,Y^4
\nonumber \\ && 
+{303 \over 224}\,Y^5
+{10661897 \over 4718592}\,Y^6 \,.
\end{eqnarray}
When $Q=2$, the errors induced by the above expansion are found to be 
$\sim 0.1\,\%$. This is accurate enough for our purpose 
as long as we allow the SNR loss, $ds^2_{\rm max}$, of a few percent. 

The fact that the conformal factor depends only on the coordinate $Y$
is important. It allows us to apply the same tiling method
developed in~\cite{NTTS}, which we recapitulate in Appendix~\ref{app:TM}.
In Fig.~\ref{fig:umeta1}, we show the tiling of the template space 
in the $(X,\,Y)$ coordinates. The tiling in the original coordinates
 $(f_c,\,Q)$ is shown in Fig.~\ref{fig:umeta2}.

%%%%%%%%%%%%%%%%%%%%%%%%%%%%%%%%%%%
%%%%%%%%%%%%%%%%%%%%%%%%%%%%%%%%%%%
\section{Test with TAMA noise spectrum}\label{sec:TN}
%%%%%%%%%%%%%%%%%%%%%%%%%%%%%%%%%%%
%%%%%%%%%%%%%%%%%%%%%%%%%%%%%%%%%%%

The derivation of the template space metric~(\ref{eq:conformal})
is based on the assumption that the noise is white. 
This assumption may be good because a ringdown wave
will be rather narrow banded except for the case $Q\sim 2$. 
In order to confirm this, we examine the effectiveness of 
the tiling based on Eq.~(\ref{eq:conformal}) 
in the case of a colored noise.

As a model of detector's noise, 
we use a fitting curve of the one sided noise power spectrum of TAMA300, 
which is given by 
\begin{eqnarray}
S_n(|f|) &=& {3.0 \times 10 \over f^8} + {7.0 \times 10^{-16} \over f} 
+ 7.0 \times 10^{-23}\, f^{1.2} \,,
\end{eqnarray}
where the frequency $f$ is in units of Hz.
This formula of the noise spectrum is obtained 
during Data Taking 8 in 2003~\cite{Tatsumi}. 
The normalization of the overall amplitude is arbitrary.

We prepare a template bank using Eq.~(\ref{eq:conformal}).
The minimum match is required to be 0.98.
We also generate signals whose amplitudes are normalized to unity. 
Then, we perform the matched filtering and evaluate the maximum of 
the match for each signal.
If the match is always greater than 0.98, the use of the template
space metric~(\ref{eq:conformal}) is justified even for a colored noise,
at least for the TAMA DT8 noise spectrum.

\begin{figure}[ht]
\center
\epsfxsize=10cm
\hspace{-1cm}\epsfbox{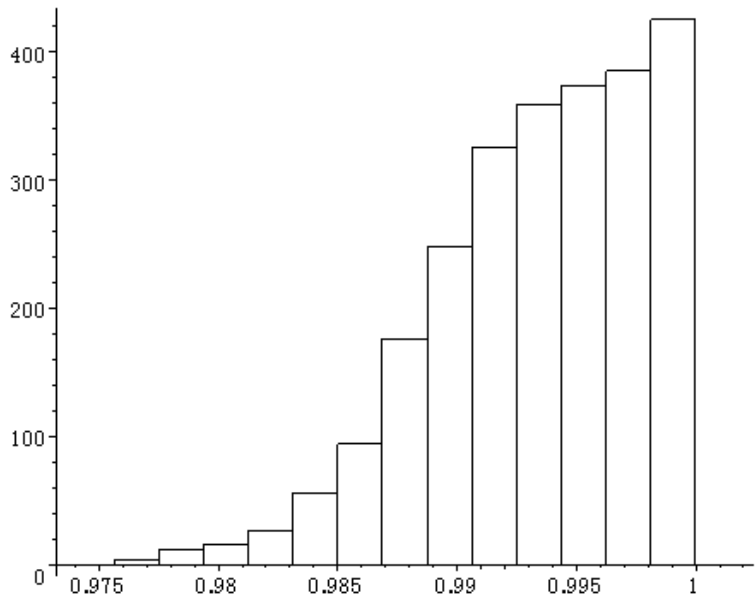}
\caption{
The number of signals 
in terms of the value of the match between the signal and templates. 
The bank of templates are determined 
assuming the minimum match 0.98 using the metric~(\ref{eq:conformal}) 
and the tiling method in \cite{NTTS}. 
The mean value of the match is $0.993$.
}
\label{fig:effnum}
\end{figure}

Here we generated 2500 signals randomly with uniform
 probability in the range of the parameters
$1.0 \times 10^2$Hz $\leq f_c \leq 2.5 \times 10^3$Hz,
$2.0 \leq Q \leq 33.3$ and $0 \leq \phi_0 \leq \pi$. 
In Fig.~\ref{fig:effnum}, we plot the number of signals 
for each bin of the match. We see that the most of the signals 
are detected with the SNR loss less than $2\%$.
Thus, the template spacing, constructed analytically 
under the assumption of the white noise, turns out to be valid even 
in the case of the colored TAMA noise spectrum. 

%%%%%%%%%%%%%%%%%%%%%%%%%%%%%%%%%%%
%%%%%%%%%%%%%%%%%%%%%%%%%%%%%%%%%%%
\section{Discussion}\label{sec:Dis}
%%%%%%%%%%%%%%%%%%%%%%%%%%%%%%%%%%%
%%%%%%%%%%%%%%%%%%%%%%%%%%%%%%%%%%%

The detection of ringdown waves is a direct confirmation of 
the existence of a black hole. 
The ringdown waves have damped sinusoidal waveforms that
reflect the mass and spin of a black hole.
In our previous paper~\cite{NTTS}, we proposed 
an efficient method for tiling the templates 
for matched filtering in the 2-dimensional $\{f_c,\,Q\}$ space.
However, it relied on the template space metric 
for signals with the initial phase $\phi_0=0$.
In this paper, we took account of the initial phase dependence
and developed a new, improved method to search 
for the ringdown waves.

Since the template metric depends on the initial phase,
we first determined the inner envelope curve of all the
contours of a fixed maximum distance $ds^2_{\rm max}$
between signals and templates on the $(f_c,\,Q)$-plane
for all possible values of the initial phase. 
We then constructed an ellipse that can cover the region
inside the inner envelope as much as possible.
Finally, we applied our tiling method proposed in~\cite{NTTS}
to obtain an improved, reliable tiling of the template space.

Another change from our previous work is the difference in the
definition of the match. In our previous work, we defined the match
as the square of the SNR, $\rho^2$, while here we used the SNR as the 
match. So, for the same fixed maximum distance,
the number of templates needed to cover the
template space turns out to be smaller than what we obtained 
previously, if we ignore the initial phase dependence.
 As a result, the actual number of templates
to cover the template space did not increase much from
what was suggested in~\cite{NTTS}.

We also examined the validity of our tiling method
in the case of a colored noise spectrum by a 
Monte Carlo simulation. 
As a model of realistic noise power spectrum, we used 
a fitting curve of the noise power spectrum of TAMA300 
during DT8 in 2003. 
For the pre-assigned maximum allowable SNR loss of $2\%$,
we found that only a few out of 2500 signals had
the SNR loss larger than $2\%$.
This means that our template spacing is effective even 
in the case of colored noise. 

Since the real data includes the non-stationary, non-Gaussian noise,
it is necessary to check the effectiveness of this method in a
realistic situation. Using the real data of TAMA300, 
this new template spacing is now being tested by 
Tsunesada et al.~\cite{Tsune}. 
They find a lot of fake events due to the
non-stationary, non-Gaussian noise. 
Particularly, the detector has many noise sources that can
produce fake ringdown wave signals.
Apparently,
we need to develop a method to remove these fake events
without losing real gravitational ringdown wave signals.
Perhaps, the best way is to perform a coincidence analysis
if we have plural detectors.
We plan to study methods of coincidence or coherent analyses 
by using several detectors in the future.

%===========================%
\acknowledgments
%===========================%

We would like to thank N. Kanda and Y. Tsunesada for invaluable 
discussions and advice. 
We are also grateful to D. Tatsumi for useful comments. 
HN was supported by the Japan Society for the Promotion of Science 
for Young Scientists, No.~5919. 
This work was supported in part by the Grant-in-Aid for 
Scientific Research on Priority Areas (415) of the Ministry of Education, 
Culture, Sports, Science and Technology of Japan, 
and in part by Grant-in-Aid 
for Scientific Research Nos.~14047214 and 12640269.

%======================================%
%<<<<<<<<<<<<< APPENDIX >>>>>>>>>>>>>>>%
%======================================%

\begin{appendix}

%%%%%%%%%%%%%%%%%%%%%%%%%%%%%%%%%%%
\section{Efficient Tiling method}\label{app:TM}
%%%%%%%%%%%%%%%%%%%%%%%%%%%%%%%%%%%

In Subsec.~\ref{subsec:CFM}, 
we have derived the simple, conformally flat metric~(\ref{eq:ds2coN}) 
for the template space. Here, using this metric, we formulate 
a tiling algorithm which is not only efficient but also quite simple.

%%%%%%%%%%%%%%%%%%%%%%%%%%%%%%%%%%%
\subsection{Basis}
%%%%%%%%%%%%%%%%%%%%%%%%%%%%%%%%%%%

To develop such a method, we note the following. 
Because of the conformal flatness, the contour of the fixed maximum 
distance $ds^2=ds^2_{\rm max}$ centered at a point on the 
$(X,Y)$-plane is a circle for sufficiently small $ds^2_{\rm max}$. 
Furthermore, along a line of $Y=$constant, $\Omega(Y)$ is constant. 
Thus, choosing first an appropriate $Y=$constant line, 
say $Y=q_1$, we may place circles of the same radius with their 
centers located along the line $Y=q_1$ to cover a region 
surrounding that line. Then, if we find an algorithm to 
place circles along the $Y=q_1$ line and another algorithm to 
choose the next $Y=$constant line, say $Y=q_2$, to be covered in an 
appropriate way, we can repeat this tiling procedure to 
cover the whole template space. 

Let us assume that the template space to be tiled is 
a rectangle given by $F_{\rm min}\leq F\leq F_{\rm max}$ 
and $Q_{\rm min}\leq Q\leq Q_{\rm max}$. In the $(X,Y)$ 
coordinates, this rectangle is mapped to the region 
bounded by the two $Y=$constant lines 
corresponding to $Q=Q_{\rm min}$ and $Q=Q_{\rm max}$, 
which we denote by $Y=Y_0$ and $Y_{\rm M}$, respectively, 
and the two lines $X=z_{\rm min}(Y)$ and $X=z_{\rm max}(Y)$ 
corresponding to $F=F_{\rm min}$ and $F=F_{\rm max}$, respectively. 
Note that $Y_0>Y_{\rm M}$ since large $Y$ corresponds to small $Q$. 

\begin{figure}[ht]
\begin{center}
\epsfxsize=8cm
\hspace{-1cm}\epsfbox{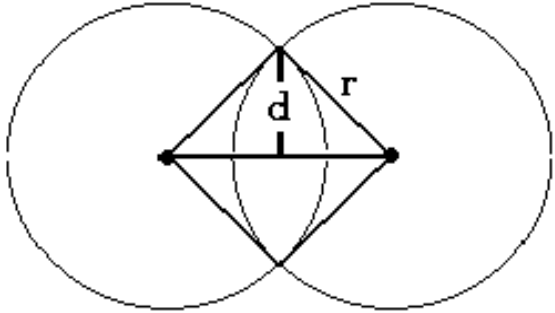}
\caption{Definition of $d$ and $r$.}
\label{fig:umo12}
\end{center}
\end{figure}

First, we construct a method to determine 
the spacing of the circles along each $Y=$constant line. 
Let us consider the line $Y=q$ and 
place two circles with radius $r$ 
centered at $(p,\,q)$ and $(p+\Delta p,\,q)$,
\begin{eqnarray}
(X-p)^2 + (Y-q)^2 = r^2 \,, 
\quad
(X-p-\Delta p)^2 + (Y-q)^2 = r^2 \,.
\end{eqnarray}
We assume $\Delta p<2r$ so that the two circles intersect at 
the two points, $\left(p+{\Delta p/2}\,,\,q\pm d(r;\,p)\right)$,
where $d(r;\,p)$ is the distance to each intersecting point
from the line $Y=q$ (see Fig.\ref{fig:umo12}), given by 
\begin{eqnarray}
d(r;\,\Delta p) = \sqrt{r^2-{\Delta p^2 \over 4}} \,. 
\end{eqnarray}
Our purpose is to tile the template space by the smallest 
possible number of filters. In order to do so, we choose the 
parameter $p$ in such a way that the area defined by 
$S=\Delta p\,d(r;\,\Delta p)$ is maximized, i.e., 
\begin{eqnarray}
\Delta p &=& \sqrt{2}\,r \,, \\
d &=& {r \over \sqrt{2}} \,.
\end{eqnarray}
The radius $r$ is determined by 
the value of $ds_{\rm max}^2$ and $q$ as 
\begin{eqnarray}
r^2={ds_{\rm max}^2 \over \Omega(q)}\,.
\label{eq:rdet}
\end{eqnarray}
In this way, we tile the region that covers the line $Y=q$. 

\begin{figure}[ht]
\begin{center}
\epsfxsize=8cm
\hspace{-1cm}\epsfbox{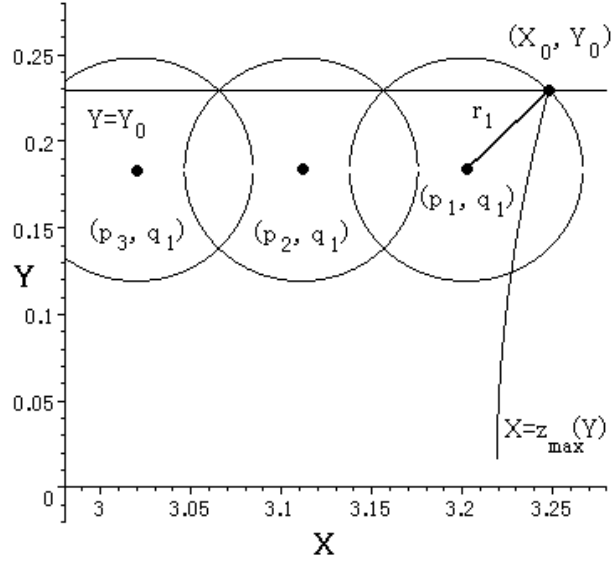}
\caption{Choosing the first line on the $(X,Y)$-plane.
}
\label{fig:umo1}
\end{center}
\end{figure}

To choose the first line to be covered, we start from 
the point on the $(X,Y)$-plane
corresponding to $(F,Q)=(F_{\rm max},\,Q_{\rm min})$,
that is, $(X,Y)=(X_0,Y_0)$ where $X_0=z_{\rm max}(Y_0)$. 
Then we choose the first line $Y=q_1$ and the radius $r_1$
so that the point $(X_0,Y_0)$ is just on the edge of the first
circle and an intersecting point of the first and second circles
lies on the line $Y=Y_0$ as Fig.~\ref{fig:umo1}. 
This is achieved if the center of the first circle is at 
\begin{eqnarray}
(p_1,\,q_1)= 
\left({X_0-r_1/\sqrt{2}\,,\,Y_0-r_1/\sqrt{2}}\right)\,,
\end{eqnarray}
with the radius determined by solving Eq.~(\ref{eq:rdet}),
which reads in the present case,
\begin{eqnarray}
ds_{\rm max}^2=r_1^2\,\Omega(Y_0-r_1/\sqrt{2})\,.
\label{eq:r1eq}
\end{eqnarray}

Then the center of the $n$-th circle is at
\begin{eqnarray}
(p_n,\,q_1)=
\left({X_0-(2n-1)\,r_1/\sqrt{2}\,,\,Y_0-r_1/\sqrt{2}}\right)\,,
\end{eqnarray}
and the number of circles needed to cover the line $Y=q_1$
is given by
\begin{eqnarray}
N_1=\left[{L_1\over \sqrt{2}\,r_1}\right],
\label{eq:N1}
\end{eqnarray}
where $[x]$ denotes the maximum integer smaller than $x+1$
and $L_1$ is the coordinate length of $X$ to be covered,
i.e., 
\begin{eqnarray}
L_1 &=& z_{\rm max}(Y_0)-z_{\rm min}(Y_0) 
\nonumber \\ 
&=& F_{\rm max} - F_{\rm min} 
\nonumber \\ &=:& L \,.
\end{eqnarray}

\begin{figure}[ht]
\begin{center}
\epsfxsize=8cm
\hspace{-1cm}\epsfbox{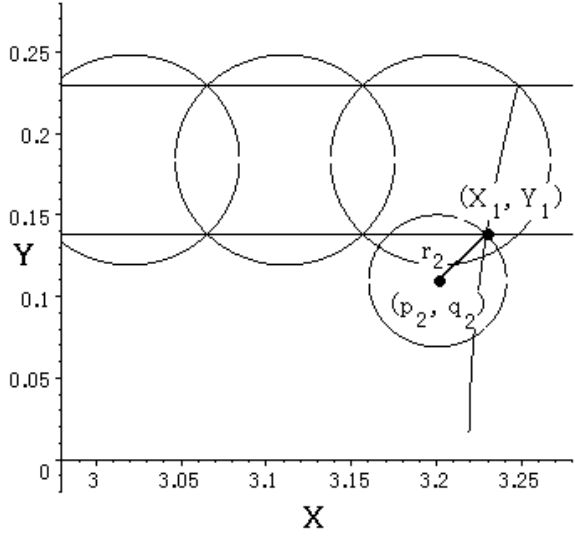}
\caption{Covering of the second line.
}
\label{fig:umo3}
\end{center}
\end{figure}

Once the covering of the first line is done, 
the second $Y=$constant line is chosen as follows.
Let $Y_1=Y_0-\sqrt{2}r_1$ and let $(X_1,Y_1)$ be the intersecting 
point of the lines $Y=Y_1$ and $X=z_{\rm max}(Y)$, 
i.e., $(X_1,Y_1)=(z_{\rm max}(Y_1),Y_1)$.
We choose this point as the starting point for the covering
of the second line as Fig.~\ref{fig:umo3}. 
That is, the center of the
first circle $(p_2,q_2)$ on the second line $Y=q_2$ is 
\begin{eqnarray}
(p_2,\,q_2)=
\left({X_1-r_2/\sqrt{2}\,,\,Y_1-r_2/\sqrt{2}}\right)\,,
\end{eqnarray}
with the radius $r_2$ determined again by Eq.~(\ref{eq:rdet}).
Then the second line is covered by the same procedure
we took for the first line. 
We repeat this procedure until we tile the whole template
space we need to cover. 

With this tiling procedure, the total number of templates
is given as follows.
Generalizing Eq.~(\ref{eq:N1}),
the number of templates for the $i$-th $Y=$constant line 
($Y=q_i$) is
\begin{eqnarray}
N_i=\left[{L \over \sqrt{2}\,r_i}\right]\,,
\end{eqnarray}
where 
$r_i$ is determined by
\begin{eqnarray}
ds_{\rm max}^2=r_i^2\,\Omega(Y_{i-1}-r_i/\sqrt{2})\,.
\end{eqnarray}
The number of $Y=$constant lines necessary to cover
the template space is determined by the minimum integer
$\nu$ that satisfies the inequality,
\begin{eqnarray}
\sum_{i=1}^{\nu}\sqrt{2}\,r_i \geq Y_0-Y_{\rm M}\,.
\end{eqnarray}
And the total number of templates is 
\begin{eqnarray}
{\cal N} &=& \sum_{i=1}^{\nu} N_i \,.
\end{eqnarray}

%%%%%%%%%%%%%%%%%%%%%%%%%%%%%%%%%%%
\subsection{Application}
%%%%%%%%%%%%%%%%%%%%%%%%%%%%%%%%%%%

Let us apply the method developed in the previous
subsection to the case of the parameter space $(f_c,\,Q)$ 
which has the range, 
\begin{eqnarray}
1. \times 10^2{\rm Hz} \leq &f_c & \leq 2.5 \times 10^3{\rm Hz} \,, 
\nonumber \\ 
2.0 \leq &Q& \leq 20.0 \,.
\end{eqnarray}
We set $ds^2_{\rm max}=0.02$. This choice is made 
in order to make the SNR loss to be $\sim 3\,\%$ 
in the presence of colored noise as discussed
in Sec.~\ref{sec:TN}. 
Using Eq.(\ref{eq:defdF}), we set
$F=\ln(f_c/100{\rm Hz})$. 
The above range corresponds to 
\begin{eqnarray}
3.12 \times 10^{-4} \leq &X& \leq 3.25 \,, \nonumber \\ 
2.48 \times 10^{-2} \leq &Y& \leq 2.29 \times 10^{-1} \,.
\end{eqnarray}
in the $(X,\,Y)$-space. (See Table~\ref{table:trans}.)
\begin{table}[ht]
 \begin{center}
\renewcommand{\arraystretch}{1.8}
  \begin{tabular}{|c|c|}
\hline
\hspace{5mm}$(f_c,\,Q)$\hspace{5mm} & \hspace{25mm}$(X,\,Y)$\hspace{25mm}  \\ 
\hline
1.0 $\times 10^2$Hz,\,2.0 & 3.030840555 $\times 10^{-2}$,\,2.293688750 $\times 10^{-1}$ \\ 
\hline
1.0 $\times 10^2$Hz,\,20.0 & 3.124023844 $\times 10^{-4}$,\,2.483796010 $\times 10^{-2}$ \\
\hline
1.0 $\times 10^3$Hz,\,2.0 & 2.331909728,\,2.293688750 $\times 10^{-1}$ \\ 
\hline
1.0 $\times 10^3$Hz,\,20.0 & 2.302897494,\,2.483796010 $\times 10^{-2}$ \\ 
\hline
2.5 $\times 10^3$Hz,\,2.0 & 3.247808978,\,2.293688750 $\times 10^{-1}$ \\ 
\hline
2.5 $\times 10^3$Hz,\,20.0 & 3.219188225,\,2.483796010 $\times 10^{-2}$ \\ 
\hline
  \end{tabular}
 \end{center} 
\caption{The relation between the two coordinates $(f_c,\,Q)$ and $(X,\,Y)$.}
\label{table:trans}
\end{table}

Following the procedure described in the previous subsection,
we choose the starting point $(X_0,\,Y_0)$ which 
corresponds to $(f_c,\,Q)=(2.5 \times 10^3{\rm Hz},2.0)$. 
The radius $r_1$ is determined by Eq.~(\ref{eq:r1eq}), or 
\begin{eqnarray}
0.02=r_1^2\,\Omega(Y_0-r_1/\sqrt{2})\,.
\end{eqnarray}
Then all the parameters for our tiling method are determined. 
In Table~\ref{table:param}, we summarize the 
tiling parameters.
\begin{table}[ht]
 \begin{center}
\renewcommand{\arraystretch}{1.8}
  \begin{tabular}{|c||c|c|c|c|}
\hline
$i$ & $(p_i,\,q_i)$ & $(X_i,\,Y_i)$ & $r_i$ & $N_i$ \\ 
\hline
0 & & 3.247808978,\, 2.293688750 $\times 10^{-1}$ & &  \\ 
\hline
1 & 3.202211769,\, 1.837716656 $\times 10^{-1}$ & 3.229142859,\, 1.381744561 $\times 10^{-1}$ 
& 6.448419198 $\times 10^{-2}$ & 36 \\
\hline
2 & 3.200527260,\, 1.095588566 $\times 10^{-1}$ & 3.222295278,\, 8.094325710 $\times 10^{-2}$ 
& 4.046856890 $\times 10^{-2}$ & 57 \\ 
\hline
3 & 3.205093210,\, 6.374118861 $\times 10^{-2}$ & 3.219985276,\, 4.653912012 $\times 10^{-2}$ 
& 2.432739855 $\times 10^{-2}$ & 94 \\ 
\hline
4 & 3.209939255,\, 3.649309938 $\times 10^{-2}$ & 3.219230312,\, 2.644707864 $\times 10^{-2}$ 
& 1.420721877 $\times 10^{-2}$ & 161 \\ 
\hline
5 & 3.213469023,\, 2.068578968 $\times 10^{-2}$ & & 8.147692989 $\times 10^{-3}$ & 280 \\ 
\hline
  \end{tabular}
 \end{center}
\caption{The parameters for the tiling of the template space are summarized. 
$(X_i,\,Y_i)$ denotes the starting point of the $i$-th template spacing
along the line $y=q_i$. 
$(p_i,\,q_i)$ denotes the center of the first circle along the $i$-th line
$y=q_i$, $r_i$ is the radius of the circle, and $N_i$ is
the number of circles necessary to cover the $i$-th line.
}
\label{table:param}
\end{table}

It is noted that the number $\nu$ of the $Y=$constant lines is 
very small, 
\begin{eqnarray}
\nu=5 \,.
\end{eqnarray}
The total number of templates is calculated to be ${\cal N}=628$ 
(${\cal N}=1121$ for $Q_{\rm max}=59.4$). 
In Table~\ref{table:NofT}, we summarize the total number of templates ${\cal N}$ 
with the range of the quality factor between $Q_{\rm min}=2.0$ 
corresponding to $a=0$ and various maximum $Q_{\rm max}^{(i)}$. 
Here $i$ means the number of $Y=$constant lines necessary 
to cover the template space, and the maximum quality factor 
is determined by this number. 
\begin{table}[ht]
 \begin{center}
\renewcommand{\arraystretch}{1.8}
  \begin{tabular}{|c|c|c|}
\hline
\hspace{5mm}$i$\hspace{5mm} & \hspace{20mm}$Q_{\rm max}^{(i)} (a)$\hspace{20mm} & 
\hspace{5mm}${\cal N}$\hspace{5mm}  \\ 
\hline
$1$ & 3.461857533 (0.7045454540) & 36 \\ 
\hline
$2$ & 6.033964925 (0.9140427132) & 93 \\
\hline
$3$ & 10.60831054 (0.9754675053) & 187 \\ 
\hline
$4$ & 18.77484416 (0.9931010039) & 348 \\ 
\hline
$5$ & 33.37367768 (0.9980786204) & 628 \\ 
\hline
$6$ & 59.48251985 (0.9994680533) & 1121 \\ 
\hline
$7$ & 106.1826585 (0.9998532378) & 1995 \\ 
\hline
  \end{tabular}
 \end{center} 
\caption{The maximum range of the quality factor and the number of templates.}
\label{table:NofT}
\end{table}

\begin{figure}[ht]
\center
\epsfxsize=8cm
\hspace{-1cm}\epsfbox{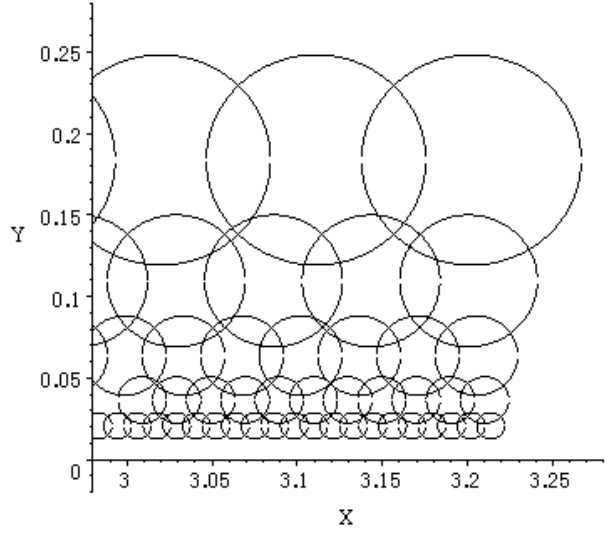}
\caption{A part of the tiling of the template space
in the $(X,Y)$ coordinates. 
Templates are taken at the centers of the circles. 
}
\label{fig:umeta1}
\end{figure}

\begin{figure}[ht]
\center
\epsfxsize=10cm
\hspace{-1cm}\epsfbox{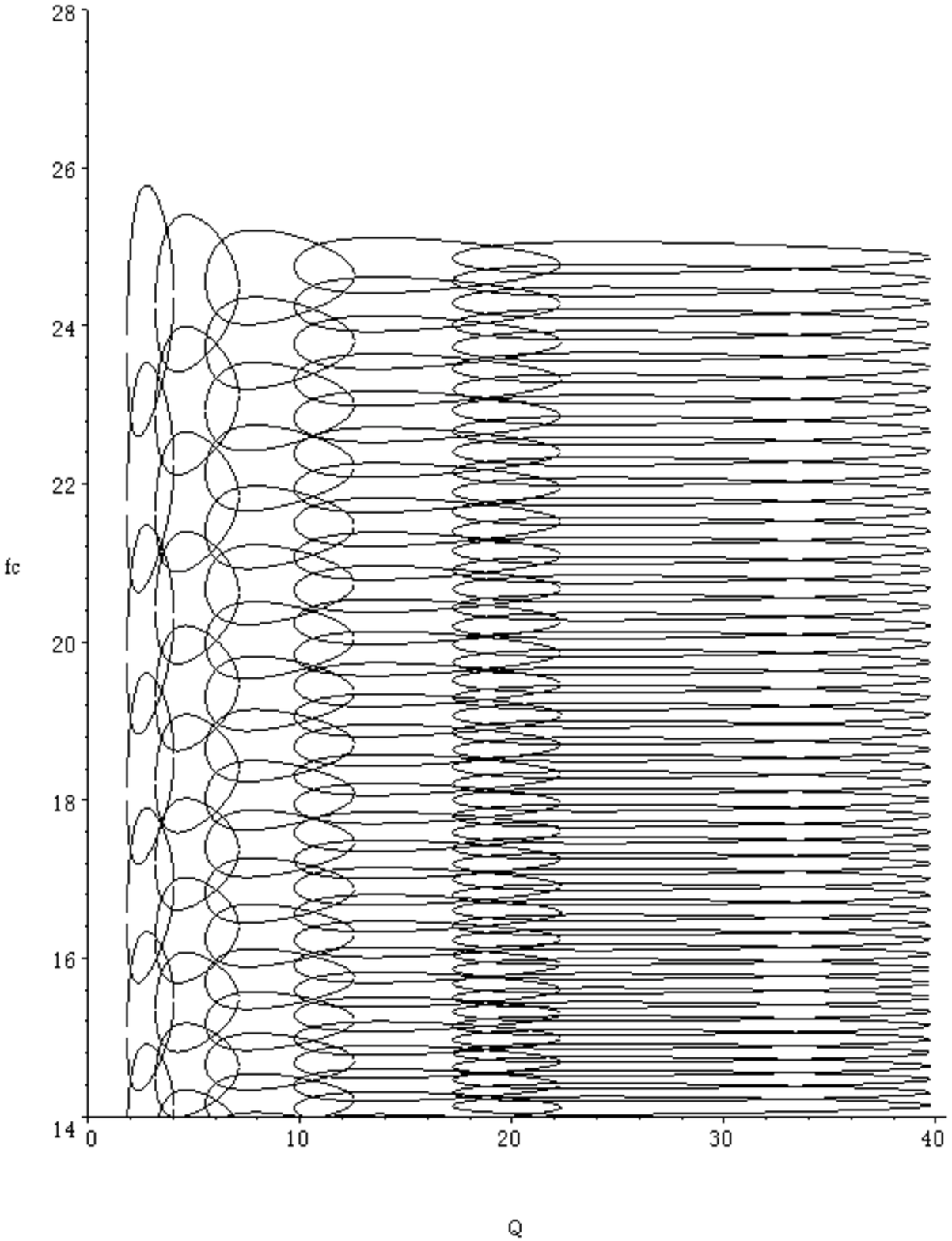}
\caption{A part of the tiling of the template space 
in the original coordinates $(f_c,\,Q)$ with $f_c$ measured in 
units of $100{\rm Hz}$.
Note that the contour of $ds^2_{\rm max}=0.02$ 
for each template is warped. 
}
\label{fig:umeta2}
\end{figure}

In Fig.~\ref{fig:umeta1}, we show the tiling of the template space 
in the $(X,\,Y)$ coordinates. 
The tiling of the template space in the original coordinates $(f_c,\,Q)$
is shown in Fig.~\ref{fig:umeta2}.

%%%%%%%%%%%%%%%%%%%%%%%%%%%%%%%%%%%
\section{Parameter estimation errors}\label{app:PEE}
%%%%%%%%%%%%%%%%%%%%%%%%%%%%%%%%%%%

Here we summarize the method to calculate the parameter estimation errors 
for gravitational ringdown waves. 
As a signal, we use the ringdown waveforms in Eq.~(\ref{eq:NGwave}) 
with the amplitude $A$, 
\begin{eqnarray}
H(A,\,f_c,\,Q,\,t_0,\,\phi_0;\,f) = A\,\hat{h}(\phi_0;\,f_c,\,Q,\,t_0;\,f) 
\,,
\end{eqnarray}
and the following calculation is done in the frequency domain. 
The parameter estimation errors are calculated by using Fisher 
information matrix 
\begin{eqnarray}
\Gamma_{\mu \nu} = \left(H_{,\mu}(A,\,f_c,\,Q,\,t_0,\,\phi_0),
H_{,\nu}(A,\,f_c,\,Q,\,t_0,\,\phi_0)\right) \,,
\end{eqnarray}
where $\mu$ and $\nu$ are $\{\ln A,\,t_0,\,\phi_0,\,f_c,\,Q\}$. 
In practice, we consider $t_0 \rightarrow f_0 t_0 $ and 
$f_c \rightarrow f_c/f_0$ 
with some frequency $f_0$ in the computational calculation. 
The root-mean-square errors $\sigma_{\mu}$ of parameters are 
obtained by 
\begin{eqnarray}
\sigma_{\mu} = \sqrt{\Sigma^{\mu \mu}} \,,
\end{eqnarray}
where ${\bf \Sigma}={\bf \Gamma}^{-1}$, and 
the parameter correlation coefficients $c^{\mu \nu}$ between 
two parameters are 
\begin{eqnarray}
c^{\mu \nu} = {\Sigma^{\mu \nu} \over \sqrt{\Sigma^{\mu \mu}\Sigma^{\nu \nu}}} \,.
\end{eqnarray}

In order to derive the Fisher information matrix, 
first we calculate $H_{,\mu}$ formally as 
\begin{eqnarray}
H_{, \ln A}(A,\,f_c,\,Q,\,t_0,\,\phi_0;\,t) 
&=& A {1 \over \sqrt{N(\phi_0;\,f_c,\,Q,\,t_0)}} 
\tilde{h}(f_c,\,Q,\,t_0,\,\phi_0;\,f) \,, 
\nonumber \\ 
H_{, t_0}(A,\,f_c,\,Q,\,t_0,\,\phi_0;\,t) 
&=& A {1 \over \sqrt{N(\phi_0;\,f_c,\,Q,\,t_0)}} 
\tilde{h}_{, t_0}(f_c,\,Q,\,t_0,\,\phi_0;\,f) \,, 
\nonumber \\ 
H_{, \alpha}(A,\,f_c,\,Q,\,t_0,\,\phi_0;\,t) &=& A 
\biggl[
-{1 \over N(\phi_0;\,f_c,\,Q,\,t_0)^{3/2}} 
(\tilde{h}_{, \alpha}(f_c,\,Q,\,t_0,\,\phi_0),\tilde{h}(f_c,\,Q,\,t_0,\,\phi_0)) 
\tilde{h}(f_c,\,Q,\,t_0,\,\phi_0;\,f) 
\nonumber \\ && \qquad 
+{1 \over \sqrt{N(\phi_0;\,f_c,\,Q,\,t_0)}} 
\tilde{h}_{, \alpha}(f_c,\,Q,\,t_0,\,\phi_0;\,f) 
\biggr] \,,
\end{eqnarray}
where $\alpha,\,\beta,\,...$ denote $\{\phi_0,\,f_c,\,Q\}$, 
and we have noted that the normalization constant $N$ 
does not depend on the initial time. 
And then, the Fisher information matrix is given by 
\begin{eqnarray}
\Gamma_{\ln A \mu} &=& A^2 \,\delta_{\ln A \mu} \,, 
\nonumber \\ 
\Gamma_{t_0 W} &=& A^2 {1 \over N(\phi_0;\,f_c,\,Q,\,t_0)} 
(\tilde{h}_{, t_0}(f_c,\,Q,\,t_0,\,\phi_0),\tilde{h}_{,W}(f_c,\,Q,\,t_0,\,\phi_0)) \,, 
\nonumber \\ 
\Gamma_{\alpha \beta} &=& A^2 
\biggl[ 
-{1 \over N(\phi_0;\,f_c,\,Q,\,t_0)^2} 
(\tilde{h}_{,\alpha}(f_c,\,Q,\,t_0,\,\phi_0),\tilde{h}(f_c,\,Q,\,t_0,\,\phi_0)) 
(\tilde{h}_{,\beta}(f_c,\,Q,\,t_0,\,\phi_0),\tilde{h}(f_c,\,Q,\,t_0,\,\phi_0)) 
\nonumber \\ && \qquad
+ {1 \over N(\phi_0;\,f_c,\,Q,\,t_0)}  (\tilde{h}_{,\alpha},\tilde{h}_{,\beta}) 
\biggr] \,,
\end{eqnarray}
where $W=\{t_0,\,\phi_0,\,f_c,\,Q\}$. 

Next, we may prepare the partial differentiation $\tilde{h}_{,W}$ of the ringdown waveforms 
to calculate the above Fisher information matrix. 
\begin{eqnarray}
\tilde{h}(f_c,\,Q,\,t_0,\,\phi_0;\,f) 
&=& {\displaystyle \frac {( \cos \phi_0\,f_c - 2\,i\,\cos \phi_0\,f\,Q 
+ 2\,f_c\,Q\,\sin \phi_0)\,Q\,e^{ 2\,i\,\pi \,f\,t_0 }}
{\pi\, ( 2\,f_c\,Q - i\,f_c - 2\,f\,Q)\,(2\,f_c\,Q + i\,f_c + 2\,f\,Q)}} 
\,, \nonumber \\ 
\tilde{h}_{,t_0}(f_c,\,Q,\,t_0,\,\phi_0;\,f) 
&=& 2\,i\,\pi \,f \,\tilde{h}(f_c,\,Q,\,t_0,\,\phi_0;\,f) 
\,,
\nonumber \\ 
\tilde{h}_{,\phi_0}(f_c,\,Q,\,t_0,\,\phi_0;\,f) 
&=& 
{2\,f_c\,Q\,\cos\phi_0-(f_c - 2\,i\,f\,Q )\,\sin\phi_0
\over 2\,f_c\,Q\,\sin\phi_0+(f_c - 2\,i\,f\,Q )\,\cos\phi_0}
\,\tilde{h}(f_c,\,Q,\,t_0,\,\phi_0;\,f)  \,,
\nonumber \\ 
\tilde{h}_{,f_c}(f_c,\,Q,\,t_0,\,\phi_0;\,f) 
&=& 
\biggl[ 
(16\,i\,f\,f_c\,Q^3-4\,(f_c^2-f^2)\,Q^2+4\,i\,f\,f_c\,Q-f_c^2)\,\cos\phi_0
\nonumber \\ && \quad
-2\,Q\,(4\,(f_c^2+f^2)\,Q^2+f_c^2)\,\sin\phi_0
\biggr]
\tilde{h}(f_c,\,Q,\,t_0,\,\phi_0;\,f) 
\nonumber \\ &&
/ \biggl[(2\,(f_c+f)\,Q + i\,f_c)\,(2\,(f_c-f)\,Q - i\,f_c)
\nonumber \\ && \quad 
\times((f_c - 2\,i\,f\,Q )\,\cos\phi_0 + 2\,f_c\,Q\,\sin\phi_0) \biggr] \,,
\nonumber \\ 
\tilde{h}_{,Q}(f_c,\,Q,\,t_0,\,\phi_0;\,f) 
&=& 
\biggl[ 
f_c ((2\,(f_c+f)\,Q + i\,f_c)\,(2\,(f_c-f)\,Q - i\,f_c)\,\cos\phi_0
\nonumber \\ && \quad 
-4\,i\,(f_c-2\,i\,f\,Q)\,f_c\,Q\,\sin\phi_0) 
\biggr] \tilde{h}(f_c,\,Q,\,t_0,\,\phi_0;\,f) 
\nonumber \\ &&
/\biggl[ 
(2\,(f_c+f)\,Q + i\,f_c)\,(2\,(f_c-f)\,Q - i\,f_c)
\nonumber \\ && \quad 
\times((f_c - 2\,i\,f\,Q )\,\cos\phi_0 + 2\,f_c\,Q\,\sin\phi_0)\,Q \biggr] 
\,.
\end{eqnarray}
In practice, by using the inner product~(\ref{eq:inner}) between two functions, 
we can obtain the root-mean-square errors 
and correlation coefficients of parameters. 

\end{appendix}

%%%%%%%%%%%%%%%%%%%%%%%%%%%%%%%%%%%%%%%%
%%%%%%%%%%%%%%%%%%%%%%%%%%%%%%%%%%%%%%%%

\end{document}